\documentclass[a4paper]{article}

\usepackage{INTERSPEECH2020}
\usepackage{xcolor}
\usepackage{amsmath}
\usepackage{enumitem}
\usepackage{multirow}
\usepackage[hidelinks]{hyperref}

\hypersetup{
	colorlinks = True,
	linkcolor=blue,
	citecolor={blue!60!black},
	urlcolor=blue
}

\newcommand{\X}{\boldsymbol{X}}
\newcommand{\Y}{\boldsymbol{Y}}
\title{Deep Self-Supervised Hierarchical Clustering for Speaker Diarization}
\name{Prachi Singh, Sriram Ganapathy\thanks{This work was funded by British Telecom India Research Center (BTIRC) project on Speech Analytics.}}
\address{
  Learning and Extraction of Acoustic Patterns (LEAP) Lab, Indian Institute of Science.}
\email{prachisingh@iisc.ac.in, sriramg@iisc.ac.in}

\begin{document}

\maketitle
\begin{abstract}
The state-of-the-art speaker diarization systems use agglomerative hierarchical clustering (AHC) which performs the clustering of previously learned neural embeddings. While the clustering approach attempts to identify speaker clusters, the AHC algorithm does not involve any further learning. In this paper, we propose a novel algorithm for hierarchical clustering which combines the speaker clustering along with a representation learning framework. The proposed approach is based on principles of self-supervised learning where the  self-supervision is derived from the clustering algorithm. The representation learning network is trained with a regularized triplet loss using the clustering solution at the current step while the clustering algorithm uses the deep embeddings from the representation learning step.  By combining the self-supervision based representation learning along with the clustering algorithm, we show that the proposed algorithm improves significantly ($29\%$ relative improvement) over the AHC algorithm with cosine similarity for a speaker diarization task on CALLHOME dataset. In addition, the proposed approach also improves over the state-of-the-art system with PLDA affinity matrix with $10\%$ relative improvement in DER.
%  (average relative improvements of $x.x$\% in DER)
\end{abstract}

\section{Introduction}
Speaker diarization, the task of determining who spoke when in a continuous multi-speaker audio stream, has received renewed interest in the recent years owing to the DIHARD evaluations \cite{ryant2019second} as well as the need for rich transcriptions in speech recognition systems \cite{settle2018end}. The main challenges to speaker diarization include short speaker turns, noise/channel distortions and overlapping speech~\cite{moattar2012review}.  

The prominent approaches to speaker diarization in the last decade have shifted to the use of embeddings from short windows of the incoming audio followed by a clustering process. Recently, the i-vector representations \cite{sell2014speaker} have been replaced with neural embeddings from a time delay neural network, called x-vectors \cite{snyder2019speaker}. The neural embeddings have the advantage of utilizing large corpora of speaker supervised data for learning the background models \cite{snyder2018x}.  While the embeddings used in diarization have advanced, the clustering approaches in many state-of-art systems use the traditional agglomerative hierarchical clustering (AHC) \cite{day1984efficient}. The improvements in the AHC algorithm targeted towards speaker diarization include pre-processing methods applied on embeddings like length normalization \cite{garcia2011analysis}, recording level principal component analysis (PCA) \cite{zhu2016online} as well as the use of probabilistic linear discriminant analysis (PLDA) based affinity matrix computation for AHC ~\cite{sell2014speaker}. However, none of these methods involve any form of learning in the clustering process.  

In this paper, we propose a joint model for clustering and representation learning using the principles of self-supervised learning. Self-supervised learning is a branch of unsupervised learning where the targets for supervision are derived from the data itself \cite{doersch2017multi,gidaris2018unsupervised}. For speech processing, self-supervised learning has been mostly attempted for phoneme and speech recognition tasks~\cite{pascual2019learning}. In this paper, we explore self-supervised learning in the AHC framework where the steps of representation learning using AHC based self-supervision and the clustering of the representations are performed in an iterative fashion. 
%With several experiments on LDC CALLHOME dataset, we illustrate the effectiveness of the proposed approach. 

\section{Related Prior Work}
Most of the speaker diarization efforts with embeddings use a clustering based algorithm for diarization. The early approach using cosine distance with K-means or spectral clustering \cite{shum2011exploiting} has been advanced with PLDA based distance metric \cite{sell2014speaker}. Narayanaswamy et. al. also proposed a metric learning paradigm for speaker diarization \cite{narayanaswamy2019designing}. A joint learning paradigm for learning embeddings from speech activity detection, overlap detection and speaker discrimination has also been attempted by Filho et. al \cite{Filho2018}. One of the drawbacks of the embedding based approach is the segmentation of the audio into fixed duration windows. Thus, speaker boundaries are constrained by the length of the windows. In order to overcome this, a re-segmentation framework using hidden Markov model and variational Bayes principles has also been explored \cite{diez2018speaker,singh2019leap}. The clustering process may impair the model optimization aimed at minimizing diarization errors. To alleviate this problem, Zhang et al.~\cite{zhang2019fully}  proposed a clustering-free diarization method.  However, this method suffers from the requirements of large amounts of supervised multi-speaker conversational data to train the model. 

In this paper, we propose a framework for joint learning of embeddings along with the clustering framework. The proposed work is inspired by Yang et. al \cite{yang2016joint}, where  joint unsupervised representation learning and clustering of images was explored. In our proposed approach, during training, clusters and representations are updated jointly. The clustering is a forward operation while the representation learning is a backward operation. 
%The motivation for this framework is that good representations are beneficial to clustering and clustering results provide self-supervisory targets for representation learning. 
With the joint learning of the two processes in a single model, we aim to obtain good representations of the audio and precise speaker clusters. 
\begin{figure}[t]
  \centering
  \includegraphics[trim={0 0 0 0},clip,width=\linewidth]{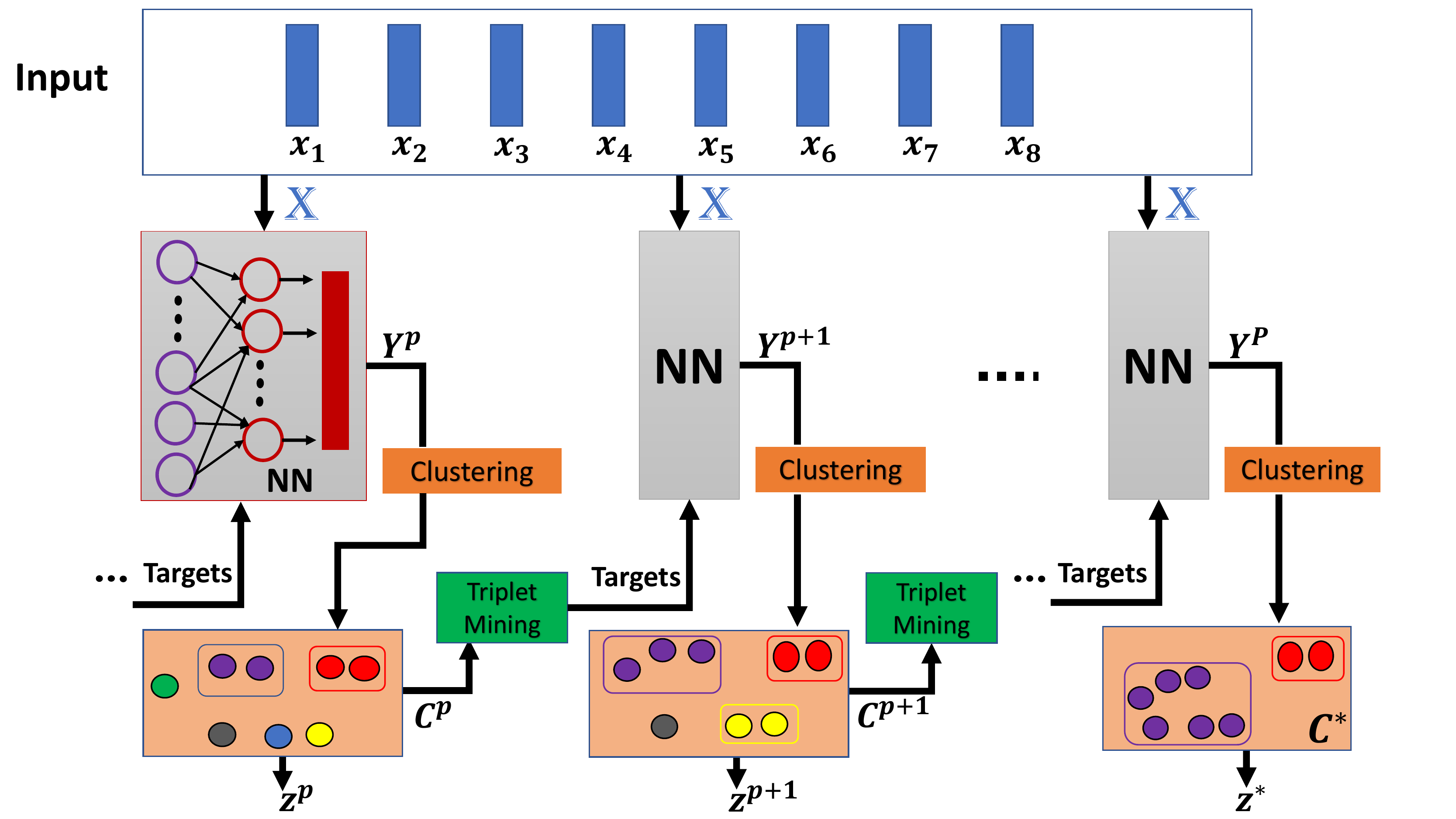}
  \caption{Block diagram of the proposed deep self-supervised clustering algorithm. }
  \label{fig:blockdiagram}
  \vspace{-6pt}
\end{figure}
\section{Proposed Approach}
The proposed work is inspired by \cite {yang2016joint}. 
The block schematic of the proposed algorithm is shown in Figure~\ref{fig:blockdiagram}. 
%The proposed algorithm iterates between learning of representations using the DNN model with cluster labels as targets and the clustering of the representations using a modified AHC algorithm.  
The inputs are the x-vector features \cite{snyder2019speaker} where each x-vector feature is extracted from a audio segment of $1.5$s duration with half overlap. More details of the time delay neural network (TDNN) based x-vector extractor are provided in Section \ref{sec:baseline}.  
%Since AHC is the most popular clustering used in Speaker Diarization we could exploit it for learning representations specific to the audio recording.
\subsection{Notations} 
The model parameters of the representation learning neural network (NN) are denoted by $\boldsymbol{\theta}$. Let $p$ denote the iteration index of the proposed algorithm.  Further, let \begin{itemize}[leftmargin=*]
    \item $\textbf{X}_r = \{x_{1},...,x_{N_r}\} \in \mathcal{R}^{D}$ denote the sequence of $N_r$ input x-vector features for the recording $r$.  
    \item $\mathbf{z}^p=\{z_1^p,...,z_{N_r}^p\}$ denote the cluster labels for this recording,  where each element takes a discrete cluster index value. \item $\textbf{Y}^p=\{y_1^p,...,y_{N_r}^p\} \in \mathcal{R}^{d}$ denote the output representations from the DNN model. 
    \item $\textbf{C}^p=\{\mathcal{C}_1^p,...,\mathcal{C}_{N^p}^p\}$ denote the cluster set where each cluster $\mathcal{C}_i^p=\{y_k^p|z_k^p=i,\forall k\in \{1,..,N_r\} \}$. Here, $N^p$ denotes the number of clusters obtained at the $p$-th iterative step.
    \item $\textbf{Q}_{\mathcal{C}_i^p}^{K_c}$ denote the set of $K_c$ nearest neighbour clusters of $\mathcal{C}_i^p$.
\item $\mathcal{A}$ denote the affinity  measure  between two clusters. 
% \item   $\mathcal{\textbf{Z}}=\{\textbf{z}^1,...,\textbf{z}^P\}$ denote the collection of labels across all $P$ iterations. 
\item $N^*$ denote the target number of clusters in the algorithm.
\end{itemize}
  
% \subsection{Input Features} X-vectors are obtained from TDNN architecture\cite{snyder2018x} as used in the \href{https://kaldi-asr.org/models/m6}{Kaldi's baseline system}. Audio is divided into short speech segments of 1.5s with 0.75s shift to extract x-vectors for each segments. These embeddings are trained with huge amount data with thousands of speakers and hence they capture speaker-specific characteristics. Hence we use these embeddings as the input to our system.
\subsection{Iterative process}\label{sec:iterative_process}
\begin{figure}[t!]
  \centering
  \includegraphics[trim={5cm 0 12cm 0},clip,width=\linewidth]{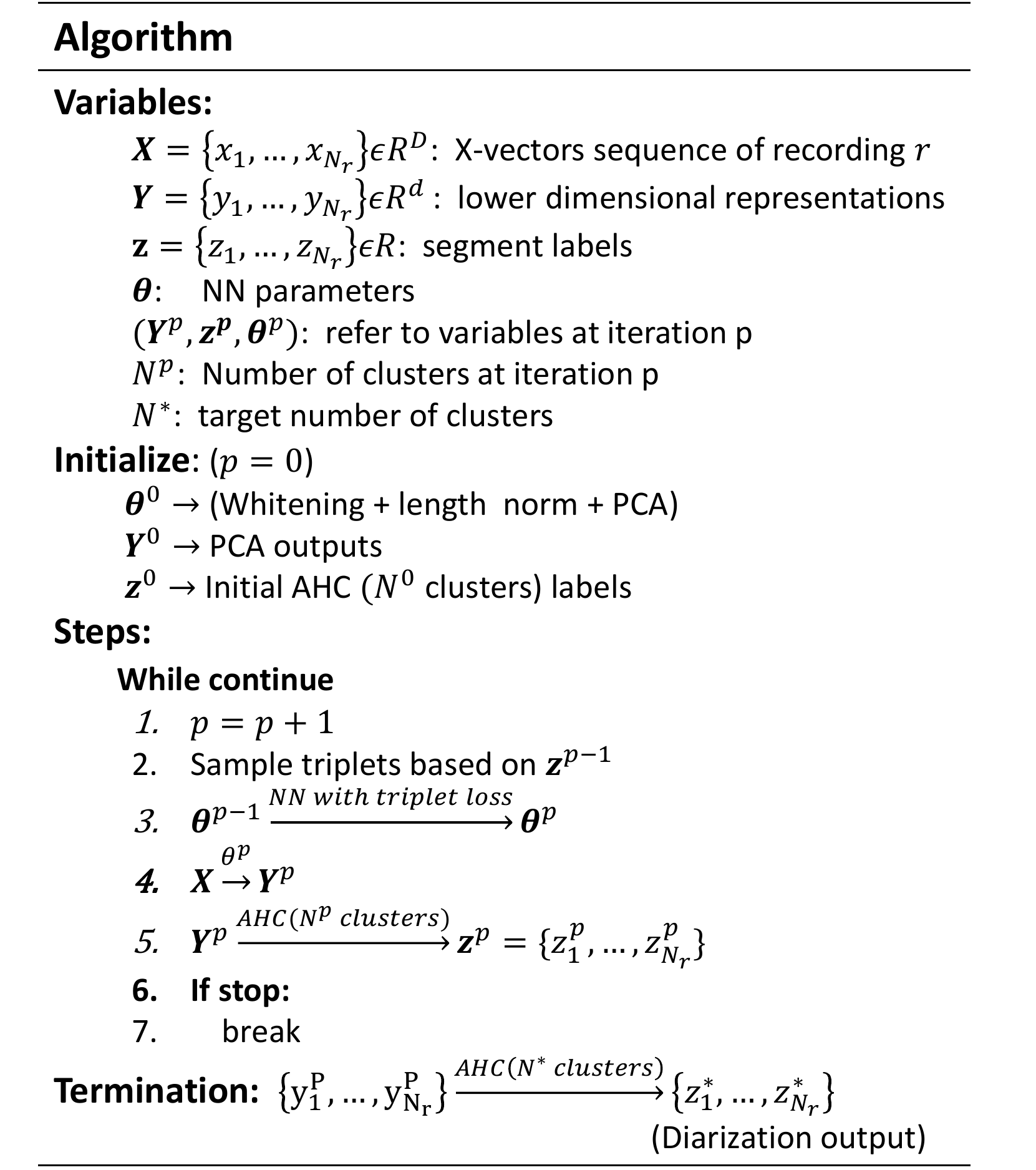}
  \caption{Algorithm for joint learning using clustering based triplet loss.}
  \label{fig:algo}
  \vspace{-10pt}
\end{figure}

The algorithm is given in Figure~\ref{fig:algo}. 
Here, we describe one step of the iterative process. At the $p$-th step, the DNN model parameters $\boldsymbol{\theta}^p$ are trained with the inputs as x-vector features $\textbf{X}$ with a triplet loss defined using the cluster label indices $\textbf{z}^{p-1}$ of the previous iteration. The DNN model is trained using several steps of gradient descent with this triplet loss. More details about the triplet mining for DNN training are given in the next section.  
Once the DNN model is trained, the last layer embeddings $\Y^p$ are derived by forward passing the x-vector inputs $\X$. Using the cluster definitions $\boldsymbol{\mathcal{C}}^{p-1}$ containing $N^{p-1}$ clusters, the initial set of clusters $\boldsymbol{\mathcal{C}}^{p}$ are formed with representations $\Y^p$ (i.e., initially $N^p=N^{p-1}$). Then, the merging of the clusters is performed based on the affinity measure between pairs of clusters. More specifically, in a standard AHC algorithm, if  $\mathcal{C}_{a}^p, \mathcal{C}_{b}^p$ are chosen as the two clusters to merge, then they satisfy,
\vspace{-4pt}
\begin{equation}
\left\{\mathcal{C}_{a}^p, \mathcal{C}_{b}^p\right\}=\underset{C_{i}^p, C_{j}^p \in \mathcal{\textbf{C}}^p, i \neq j}{\operatorname{argmax}} \mathcal{A}\left(\mathcal{C}_{i}^p, \mathcal{C}_{j}^p\right)
\label{eq1}
\end{equation}
In a modified version of the AHC merging cost, we can also incorporate discriminative term as follows,
\begin{equation}
\left\{\mathcal{C}_{a}^p, \mathcal{C}_{b}^p\right\}=\underset{C_{i}^p, C_{j}^p \in \mathcal{\textbf{C}}^p, i \neq j}{\operatorname{argmax}} \mathcal{A}\left(\mathcal{C}_{i}^p, \mathcal{C}_{j}^p\right) - \lambda \sum _{k=1}^{K_c} \mathcal{A} \left(\mathcal{C}_{ij}^{p}, Q_{\mathcal{C}_{ij}^{p}}^{K_{c}}[k]\right)
\label{eq2}
\end{equation}
where $\mathcal{C}_{ij}^{p} = \mathcal{C}_{i}^{p}\cup \mathcal{C}_{j}^{p}$ denotes the merge of clusters $i$ and $j$. Here, $Q_{\mathcal{C}_{ij}^{p}}^{K_{c}}$ denotes nearest neighbour clusters of $\mathcal{C}_{ij}^{p}$. 

While the modified cost function in Eq.~(\ref{eq2}) is more computationally expensive compared to the direct cost function in Eq.~(\ref{eq1}), the  modified cost function allows the clusters to be chosen for merging that are also distinct from other clusters in the mix. The parameter $\lambda$ is a hyper-parameter and it acts as regularization between the intra cluster affinity versus inter cluster affinity. We perform multiple merges using the criterion defined above (Eq.~(\ref{eq1})) reducing the total number of clusters in $p$-th iteration to the preset number of clusters $N^p$. 
If $N^p=N^*$, then the algorithm is stopped. Else, the iteration index is updated and the steps described above for the $p$-th iteration are repeated.  
\begin{figure*}[t]
  \centering
  \includegraphics[trim={0 3cm 0 1cm},clip,width=0.8\linewidth]{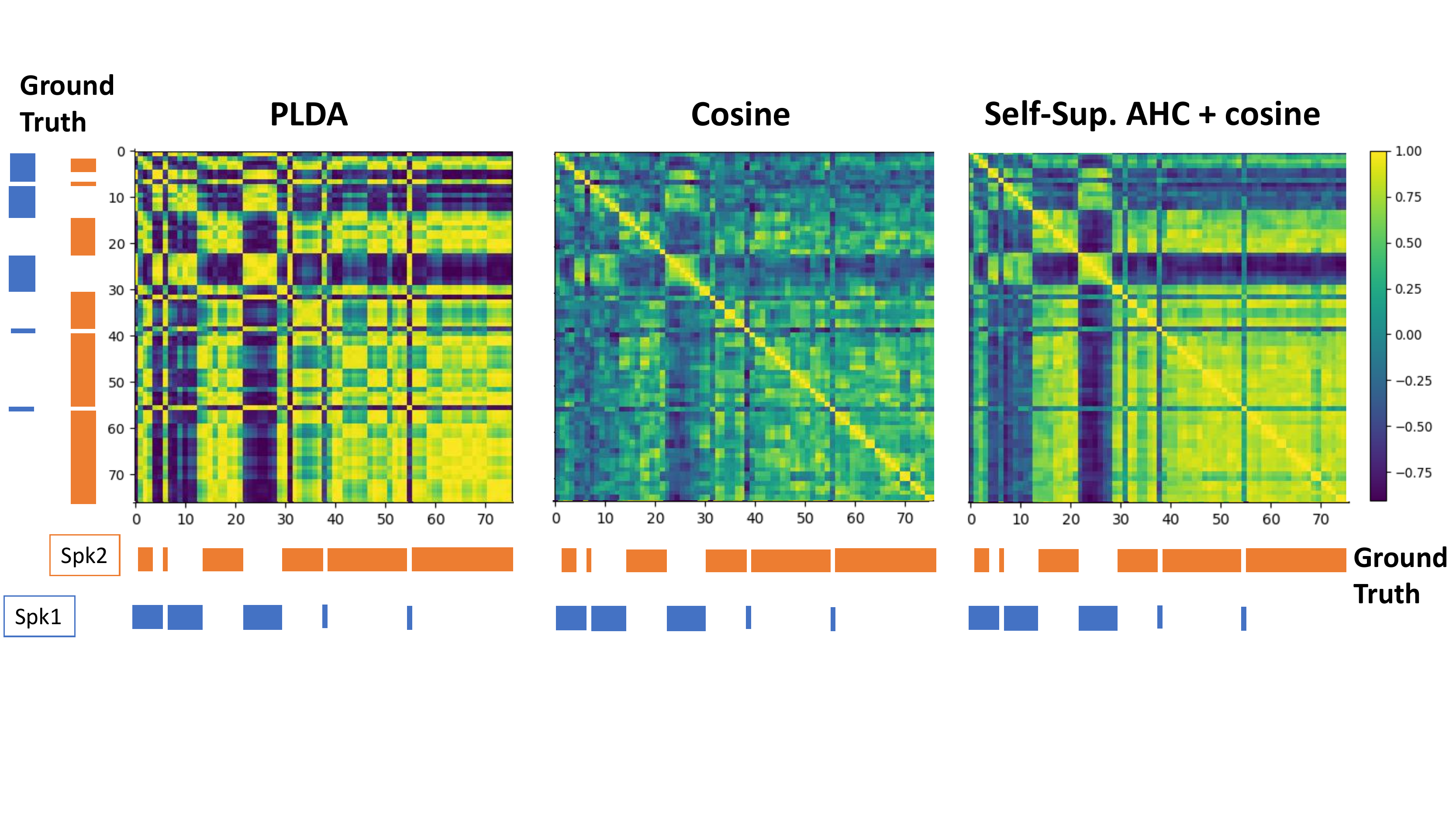}
  \vspace{-0.6cm}
  \caption{Affinity matrices using PLDA, cosine and self-supervised AHC model with cosine for a 2 speaker recording from CH dataset. Ground truth labels are plotted across time in both sides of affinity matrices for comparison.}
    \vspace{-0.4cm}
  \label{fig:affinity}
\end{figure*}
\subsection{Triplet loss for DNN Training}
For the DNN training at iteration $p$, the corresponding cluster label $z_n^p$ for each input x-vector $x_n$ is used where $n=1,..,N_r$. The DNN model  is trained by optimizing the triplet loss function in the following manner. For every cluster $\mathcal{C}_i^p$, a pair of embeddings forming the positive pair $\{y_a,y_b\}\epsilon C^{p}_i$ is selected. The negative pair in the triplet is mined using the random sampling strategy. Here, we sample one $y_c$(negative) randomly from any other cluster $\mathcal{C}_j^p$ where $j \neq i$. 
% \begin{itemize}
%     \item Random Sampling: This is the most general approach. Here, we sample one $y_c$(negative) randomly from any other cluster $\mathcal{C}_j^p$ where $j \neq i$. 
    % \item Hard Sampling: This is the more aggressive approach where we want the embeddings to be more discriminative. We sample the negative sample $y_c$ from any other cluster $\mathcal{C}_j^p$ where $j \neq i$ which has the closest (in similarity measure) to $\mathcal{C}_i^p$.
% \end{itemize}
Following the triplet mining, the model parameters of the DNN model are updated based on the following optimization criterion,
\vspace{-8pt}
\begin{equation}
\vspace{-8pt}
\boldsymbol{\theta}^p =   \underset{\boldsymbol{\theta}}{\operatorname{argmax}}\sum_{a, b, c} \big [ \mathcal{A}(y_a,y_b) - \gamma   \mathcal{A}(y_a,y_c) \big ]
\label{eq3}
\end{equation}
The above optimization is achieved using the iterative gradient descent algorithm. The hyper-parameter $\gamma$ also controls the discriminability of the model learning. 
% The representations from the model $\textbf{Y}^p$ are obtained by forward passing the input x-vectors $\textbf{X}$ through the trained DNN model with parameters   $\boldsymbol{\theta}^p$. 

\subsection{Initialization}\label{sec:3.4}
The DNN model used in our algorithm is a two layer feed-forward network. Motivated by the pre-processing steps in state-of-art diarization systems, the first layer of the DNN model is fed with x-vectors of $D$ dimensions and performs a linear transform  and a length normalization. The second layer of the DNN performs a dimensionality reduction to output representations of dimension $d$. The first layer of the DNN is initialized using a global whitening transform of the x-vectors computed using the training data. The second layer is initialized using a recording level principal component analysis (PCA) which reduces the $D$ dimensional whitened x-vectors to $d$ dimensions. The affinity measure used in this paper is based on cosine similarity measure. For initializing the clustering process, we use the AHC based clustering to $N^0$ clusters performed on the PCA components of the x-vectors. Since training is self-supervised, the initialization step has a major role in the performance. 
%clustering labels will be corrupted which will result in learning incorrect representations.
\subsection{Visualization of Affinity scores matrix}
Figure \ref{fig:affinity} shows the PLDA and cosine affinity matrices for a 2 speaker recording  from the CALLHOME dataset along with affinity matrix using cosine with self-supervised AHC.
% On the sides we have plotted ground truth time aligned labels for each speaker independently. 
The figure shows that representations have become more discriminitive in the cosine space after self-supervised training. The affinity matrix is also smoother compared to PLDA affinity matrix.

\section{Experimental Setup}

\subsection{Baseline System}\label{sec:baseline} The baseline system for CALLHOME diarization uses the Kaldi recipe\footnote{\url{https://kaldi-asr.org/models/m6}}. It involves extraction of $23$ dimensional mel-frequency cepstral coefﬁcients (MFCCs). A sliding mean normalization is applied over a $3$s window. The speech segments ($1.5$s chunks with $0.75$s overlap) are converted to $128$ dimensional neural embeddings - x-vectors \cite{snyder2018x}. For the x-vector extraction, we use a time delay neural network (TDNN) with $5$ layers followed by utterance pooling (with mean and standard deviation) and $2$ affine layers mapping to the speaker targets. The output of first affine layer following the utterance level pooling is used as the x-vector embeddings \cite{snyder2018x}. 

The backend model is based on probabilistic linear discriminant analysis (PLDA) affinity score matrix.  A Whitening transform trained on held out set is applied to the training set followed by length normalization~\cite{garcia2011analysis}. An utterance level PCA is applied before PLDA scoring for dimensionality reduction~\cite{zhu2016online}.
The clustering is performed by AHC algorithm which hierarchically clusters the segments based on the PLDA affinity matrix and merges the clusters that represent the same speaker identity. The AHC stopping criterion is determined using held-out  data. 

\subsection{Data}
We train and evaluate our approach on CALLHOME (CH) dataset. The proposed algorithm is applied independently on each recording. The CH dataset is a collection of multi-lingual telephone data containing $500$ recordings where the duration of each recording ranges from $2$-$5$ mins per file. The number of speakers in each recording vary from $2$ to $7$, with majority of the files having $2$ speakers.   The CH  dataset is equally divided into $2$ different sets CH1 and CH2 with similar distribution of number of speakers. Input to the model are x-vectors extracted for each file. For training x-vector TDNN model, we used SRE 04-08, Switchboard cellular datasets as given in the Kaldi recipe. This training set had $4,285$ speakers. 

\subsection{Self-supervised AHC modeling}
The experimental setup\footnote{\url{https://github.com/iiscleap/self_supervised_AHC}} involves following steps: 
\begin{itemize}[leftmargin=*]
    \item X-vector extraction: We use the same setup as in the baseline system to generate the x-vector features.  The initialization of the DNN model uses whitening transform and the recording level PCA from the baseline system.
    
    \item DNN model: The first layer has $128$ input and hidden nodes. The second layer has $10$ output nodes (similar to the PCA dimension used in the baseline system). The learning rate is set as $0.001$. The model is trained using the triplet loss defined in Eq.~(\ref{eq3}). The number of x-vectors per recording range from $50$-$700$. Hence, we do not split the training triplet samples into minibatches, but rather perform a full batch training with Adam optimizer~\cite{kingma2014adam}. The stopping criterion is based on the fraction of epoch training loss with initial triplet loss before the DNN training.  The DNN training is stopped when this fraction reaches a preset threshold $\eta$ (to avoid over fitting to self-supervised cluster targets).
    \item Choice of hyper-parameters: We obtained optimal $K_c$ as $1$ and fixed in all experiments. The value of $\lambda$ is kept to $0$ or $0.1$. The parameter $\eta$ is kept at $0.5$. The parameter $\gamma$ is varied between $0.2$ and $0.6$. 
    % The list and values of all parameters are given in table \ref{table:param}.
    Note that, $\lambda=0.0$ is equivalent to the original AHC based triplet mining given in Eq.~(\ref{eq1}).
  
%     \item DIHARD set:
%\end{itemize}
% \begin{table}[]
% \begin{tabular}{llllll}
% \hline
% Hyper-parameters & $K_s$ & $K_c$ & $\lambda$ & $\gamma$  & $\eta$    \\ \hline
% Value range      & 5     & 1     & \{0,0.1\} & [0.1,0.4] & 0.5 \\ \hline
% \end{tabular}
% \caption{Hyper-parameters in our approach}
% \label{table:param}
% \end{table}
\item Termination:\label{sec:termination} To estimate the final number of clusters, we stop the merging when the affinity scores between clusters are lower than the pre-fixed threshold. To find the optimum threshold, we find the DER over a range of thresholds and use the threshold which gives the best DER on the development set (similar to the baseline PLDA AHC system). For CH2 experiments, the CH1 dataset is used as the held-out set and vice versa.
\item Performance metric : Diarization error rate (DER) computed using $250$ ms collar, ignoring overlapping segments. 
\end{itemize}
% \begin{table}[t!]
% \begin{center}
% \caption{DER on CH1 subset for different initial choice of number of clusters $N^0$ and the $\gamma$ parameter. The best DER results are obtained for $\gamma=0.3$ with $10$ \% of $N_r$ as the number of initial clusters.  The results are for the oracle number of speakers (known cond.).}
% \begin{tabular}{|ccc|}
% \hline
% \textbf{$N^0$ (\% of $N_r$)}                 & \textbf{DER ($\gamma$=0.1)} & \textbf{DER($\gamma$=0.3)} \\ \hline
% 50                                            & 8.32                        & 10.58                      \\ 
% 40& 8.09                        & 11.15                      \\ 
% 30                                   & 9.38                        & 10.02                      \\ 
% 20                                            & 10.12                       & 9.29                       \\ 
% 10                                            & 11.74                       & \textbf{7.95}                       \\ \hline
% \end{tabular}
% \label{tab:clusters}
% \end{center}
% \vspace{-0.3cm}
% \end{table}

\section{Results}
%The scoring script is available with baseline setup.
We evaluate the system under two conditions - (i) using  oracle number of speakers (known $N^*$), (ii) estimating the number of speakers based on held-out set (unknown $N^*$). 
\vspace{-5pt}
%The results are based on based on different choices of hyper-parameters, different stopping criteria, different levels of training etc. These are compared with baseline system discussed in section \ref{sec:baseline}. 
% We experimented with both the hard sampling and random sampling approach. The best results were obtained for random sampling which are reported here.  
\begin{figure}[t]
  \centering
  \includegraphics[trim={2cm 2cm 2cm 2cm},clip,width=\linewidth]{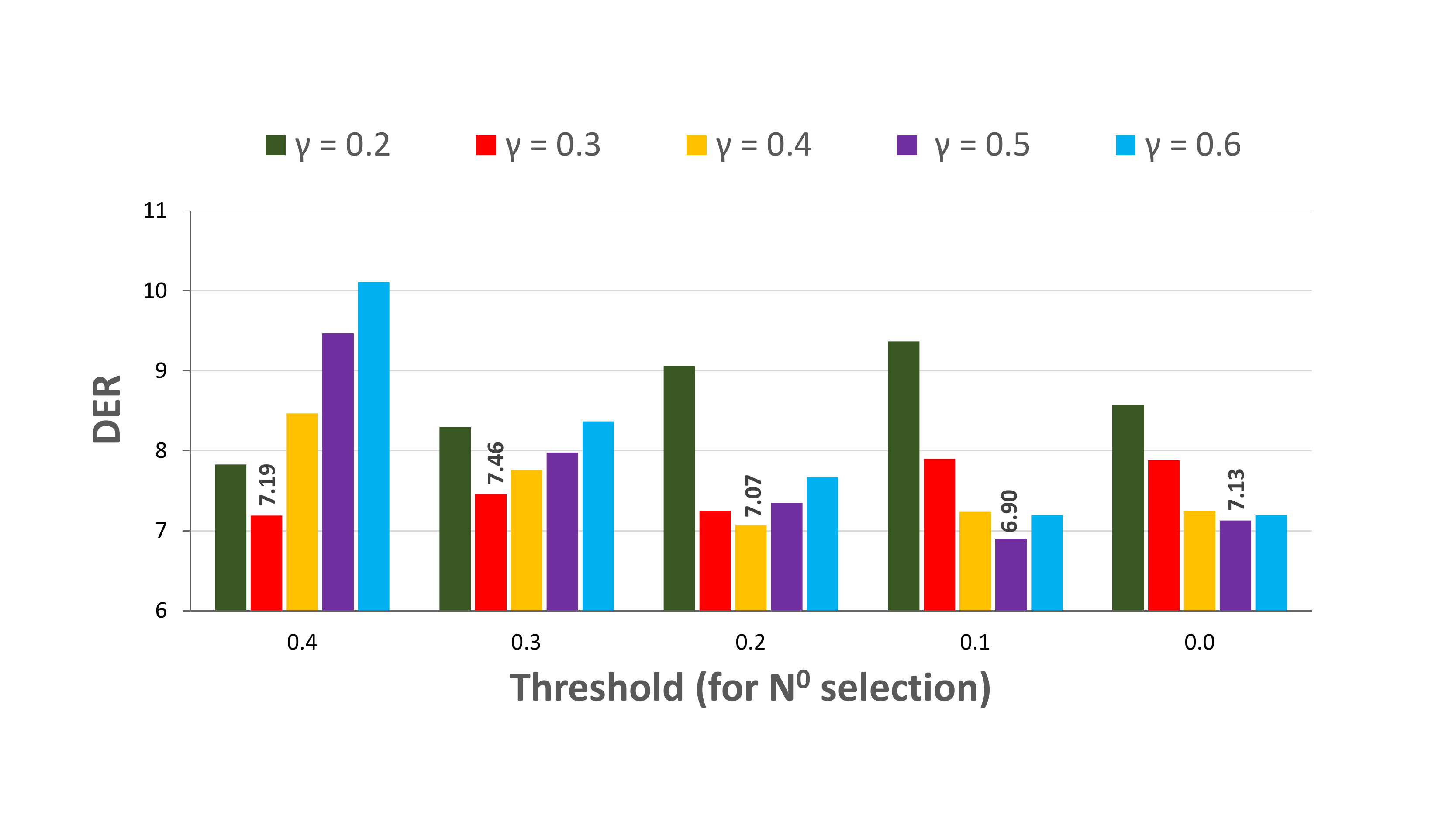}
  \caption{ Bar plot of DER vs threshold value used for selection of $N^0$ on CH1 subset for different choice of  $\gamma$ parameter with fixed $\lambda=0.1$. The best DER results (known $N^*$) are obtained for $\gamma=0.5$ with threshold=0.1.}
  \label{fig:clusters}
  \vspace{-15pt}
\end{figure}
\subsection{Choice of Initial Number of Clusters}
Figure \ref{fig:clusters} shows the impact of number of initial clusters in the self-supervised AHC algorithm. Here, the results are reported for the case when the number of speakers is known (known $N^*$). We use the threshold value as stopping criteria for AHC to decide the initial number of clusters $N^0$ in the self- supervised training. As seen in the bar plot, x-axis shows the threshold values ranging from 0.0 to 0.4. Higher threshold signifies higher number of clusters as the merging will stop early. For each threshold, we plot DER of the self- supervised AHC model for different $\gamma$ values from 0.2 to 0.6 and fixed $\lambda$ = 0.1. As we increase $\gamma$, th DER decreases till certain value and then increases again. Also, as we reduce threshold, the $N^0$ moves closer to $N^*$ and there is a significant improvement with higher $\gamma$. 
% As seen here, the number of initial clusters $N^0$ has significant impact on the DER. The more the initial number of clusters, the more the DER with a highly discriminative clustering strategy ($\gamma = 0.3$). However, when the number of initial clusters is reduced to $10$ \% of $N_r$, the use of higher value of $\gamma$ gives an improved DER.  
It is interesting to note that higher $\gamma$ results in lower DER when the number of initial clusters reduces. This may be attributed to the fact that the higher value of $\gamma$ makes the representations more discriminative across the clusters and this may result in aggressive merging of clusters thereby improving the DER. However, reducing number of clusters further starts degrading the purity of the cluster which also adversely impacts the representation learning model as well. The stability of algorithm depends on initial clustering. Hence we need to make a trade-off between number of clusters and purity. Choice of training parameters is made considering this trade-off.

\subsection{Full CALLHOME Results}
Table \ref{tab:best} shows the performance of the self-supervised AHC algorithm (SSA) compared to baseline using the known number of speakers (known $N^*$) as well as case when the number of speakers is chosen by the algorithm (unknown $N^*$). We also compared our approach with Spectral Clustering (SC) using cosine as given in \cite{wang2018speaker}. 
%The table shows DER for different choice of $\lambda$ from 0.0 to 0.2 and $\gamma$ value of 0.4 and 0.5. 
The threshold value  $th=0.1$ is chosen to generate initial number of clusters $N^0$. In the current work, we run only two iterative steps of representation learning and clustering.  
%Since the training data is small and model have few parameters to learn, we have restricted training iteration to 1 for fair comparison with different choices of $N^0$. 
%For the unknown $N^*$ case, we use the threshold based stopping criteria in AHC as described in Section \ref{sec:termination}. 

From Table~\ref{tab:best},  it can be observed that higher $\lambda$ helps in improving DER when the number of speakers is known (known $N^*$). The best DER obtained for the proposed model (SSA2) is $6.30\%$ where $\gamma=0.5$ and $\lambda=0.1$. The improvement of the self-supervised AHC  learning over the cosine similarity baseline is $29\%$. The SSA system result is also relatively better than the PLDA baseline system by about  $10\%$.  These results show that SSA algorithm provides the advantage of learning the representations in the clustering process for improved diarization.

In the case with unknown $N^*$ (number of speakers in the recording not known apriori), we find that the SSA algorithm is slightly inferior to the PLDA baseline.
%although the model improves significantly over the cosine baseline. 
We have also performed fusion of SSA with a PLDA system by averaging the affinity matrix from the final SSA algorithm and the PLDA affinity matrix. Using this combined affinity matrix for AHC improves the DER significantly for both the known $N^*$ case as well as the unknown $N^*$ case.  Overall, the fusion system (SSA+PLDA) relatively improves the PLDA baseline by $16$ \% on the known $N^*$ case and about $8$ \% in the unknown $N^*$ case. 
\begin{table}[]
\caption{Comparison of Full CH dataset DER of different systems with baseline and Spectral Clustering(SC) for both cases (i) known $N^*$ and (ii) unknown $N^*$}
\label{tab:best}
\begin{tabular}{|l|l|l|l|}
\hline
{ \textbf{System}}                                                                                                                 & { \boldsymbol{$\lambda$}} & { \textbf{\begin{tabular}[c]{@{}l@{}}DER\\ known $N^*$\end{tabular}}} & { \textbf{\begin{tabular}[c]{@{}l@{}}DER\\ unk. $N^*$\end{tabular}}} \\ \hline
{ PLDA + AHC (Baseline)}                                                                                                                 & { -}                  & {\textbf{ 7.01}}                  & { \textbf{8.00}}                     \\ 
{ Cosine + AHC}                                                                                                                          & { -}                  & { 8.86}                  & { 10.00}      \\
{ Cosine + SC~\cite{wang2018speaker}}                                                                                                                          & { -}                  & { 9.37}                  & { 11.86}\\ \hline \hline
{ }                                                                                                                                & { 0.0}                & { 6.35}                  & { \textbf{8.26}}                     \\
{ }                                                                                                                                & { 0.1}                & { 6.47}                  & { 8.33}                     \\
\multirow{-3}{*}{{\begin{tabular}[c]{@{}l@{}}Self-supervised AHC\\$[\gamma=0.4, th=0.1]$  \\(SSA1)\end{tabular}}}  & { 0.2}                & { 6.32}                  & { 8.36}                     \\ 
{ }                                                                                                                                & { 0.0}                & { 6.37}                  & { {8.51}}                     \\
{ }                                                                                                                                & { 0.1}                & { \textbf{6.30}}                  & { 8.60}                     \\
\multirow{-3}{*}{{\begin{tabular}[c]{@{}l@{}}Self-supervised AHC\\$[\gamma=0.5, th=0.1]$\\(SSA2)\end{tabular}}} & { 0.2}                & { 6.65}                  & { 8.77}                     \\ \hline \hline
{SSA1 + Baseline}                                                                                                                 & { 0.0}                & { \textbf{5.88}}                  & { \textbf{7.38}}                     \\ 
{SSA2 + Baseline}                                                                                                                 & { 0.0}                & {{ 6.10}}                  & { {7.50}}                     \\ \hline
\end{tabular}
\vspace{-10pt}
\end{table}
% \begin{table}[t!]
% \caption{Comparision of CALLHOME DER of different systems with baseline.$*$ indicates DER is evaluated using number of speakers obtained from baseline}
% \begin{tabular}{|c|c|c|c|c|}
% \hline
% \textbf{System} & \textbf{$\lambda$} & \textbf{Init.}  & \textbf{\begin{tabular}[c]{@{}l@{}}DER\\ (oracle)\end{tabular}} & \textbf{\begin{tabular}[c]{@{}l@{}}DER\\ (threshold)\end{tabular}} \\ \hline
% PLDA (Baseline)    & 0.0                & -                        & 7.01  &8.00       \\ 
% Cosine   & 0.0                & -                        & 8.86        &10 \\ 
% Sys1 [$\gamma=0.3$]          & 0.0                & PCA                      & 6.79         &8.66\\ 
% Sys1          & 0.1                & PCA                      & 6.50        & - \\ 
% Sys1          & 0.2                & PCA                      & 6.49         & -\\ 
% Sys2 [$\gamma=0.4$]          & 0.1                & PCA                      & 6.30        &8.26 \\ 
% Sys2          & 0.1                & Sys1 [$\lambda$=0.1] & 6.27       &8.20 (8.01$^*$)  \\
% Sys2$+$Baseline   & 0.1  & PCA & 5.90 & 7.38 \\
% \hline
% \end{tabular}

% \label{tab:best}
% \end{table}
\section{Summary}
We have proposed an approach to jointly learn deep representations and speaker clusters to perform diarization. This algorithm is based on principles of self-supervised learning where the self-supervision is derived from the cluster identities provided by the AHC algorithm and is implemented at the recording level. Using an iterative procedure of representation learning and clustering, we show that the proposed self supervised AHC algorithm provides improved representations and precise speaker clusters.  When the number of speakers in the recording is known, the proposed algorithm improves the baseline AHC significantly. Further, the fusion of the proposed approach with PLDA based diarization system improves the DER noticeably when the number of speakers in the given recording is unknown.  
\bibliographystyle{IEEEtran}

\bibliography{template}

\begin{thebibliography}{10}
\providecommand{\url}[1]{#1}
\csname url@samestyle\endcsname
\providecommand{\newblock}{\relax}
\providecommand{\bibinfo}[2]{#2}
\providecommand{\BIBentrySTDinterwordspacing}{\spaceskip=0pt\relax}
\providecommand{\BIBentryALTinterwordstretchfactor}{4}
\providecommand{\BIBentryALTinterwordspacing}{\spaceskip=\fontdimen2\font plus
\BIBentryALTinterwordstretchfactor\fontdimen3\font minus
  \fontdimen4\font\relax}
\providecommand{\BIBforeignlanguage}[2]{{%
\expandafter\ifx\csname l@#1\endcsname\relax
\typeout{** WARNING: IEEEtran.bst: No hyphenation pattern has been}%
\typeout{** loaded for the language `#1'. Using the pattern for}%
\typeout{** the default language instead.}%
\else
\language=\csname l@#1\endcsname
\fi
#2}}
\providecommand{\BIBdecl}{\relax}
\BIBdecl

\bibitem{ryant2019second}
N.~Ryant, K.~Church, C.~Cieri, A.~Cristia, J.~Du, and S.~Ganapathy, ``The
  second {DIHARD} diarization challenge: Dataset, task, and baselines,'' in
  \emph{Proc. of Interspeech}, 2019, pp. 978--982.

\bibitem{settle2018end}
S.~Settle, J.~Le~Roux, T.~Hori, S.~Watanabe, and J.~R. Hershey, ``End-to-end
  multi-speaker speech recognition,'' in \emph{IEEE International Conference on
  Acoustics, Speech and Signal Processing (ICASSP)}, 2018, pp. 4819--4823.

\bibitem{moattar2012review}
M.~H. Moattar and M.~M. Homayounpour, ``A review on speaker diarization systems
  and approaches,'' \emph{Speech Communication}, vol.~54, no.~10, pp.
  1065--1103, 2012.

\bibitem{sell2014speaker}
G.~Sell and D.~Garcia-Romero, ``Speaker diarization with plda i-vector scoring
  and unsupervised calibration,'' in \emph{IEEE Spoken Language Technology
  Workshop (SLT)}, 2014, pp. 413--417.

\bibitem{snyder2019speaker}
D.~Snyder, D.~Garcia-Romero, G.~Sell, A.~McCree, D.~Povey, and S.~Khudanpur,
  ``Speaker recognition for multi-speaker conversations using x-vectors,'' in
  \emph{IEEE International Conference on Acoustics, Speech and Signal
  Processing (ICASSP)}, 2019, pp. 5796--5800.

\bibitem{snyder2018x}
D.~Snyder, D.~Garcia-Romero, G.~Sell, D.~Povey, and S.~Khudanpur, ``X-vectors:
  Robust dnn embeddings for speaker recognition,'' in \emph{IEEE International
  Conference on Acoustics, Speech and Signal Processing (ICASSP)}, 2018, pp.
  5329--5333.

\bibitem{day1984efficient}
W.~H. Day and H.~Edelsbrunner, ``Efficient algorithms for agglomerative
  hierarchical clustering methods,'' \emph{Journal of classification}, vol.~1,
  no.~1, pp. 7--24, 1984.

\bibitem{garcia2011analysis}
D.~Garcia-Romero and C.~Y. Espy-Wilson, ``Analysis of i-vector length
  normalization in speaker recognition systems,'' in \emph{Twelfth annual
  conference of the international speech communication association}, 2011.

\bibitem{zhu2016online}
W.~Zhu and J.~Pelecanos, ``Online speaker diarization using adapted i-vector
  transforms,'' in \emph{IEEE International Conference on Acoustics, Speech and
  Signal Processing (ICASSP)}, 2016, pp. 5045--5049.

\bibitem{doersch2017multi}
C.~Doersch and A.~Zisserman, ``Multi-task self-supervised visual learning,'' in
  \emph{Proceedings of the IEEE International Conference on Computer Vision},
  2017, pp. 2051--2060.

\bibitem{gidaris2018unsupervised}
S.~Gidaris, P.~Singh, and N.~Komodakis, ``Unsupervised representation learning
  by predicting image rotations,'' in \emph{ICLR}, 2018.

\bibitem{pascual2019learning}
S.~Pascual, M.~Ravanelli, J.~Serr{\`a}, A.~Bonafonte, and Y.~Bengio, ``Learning
  problem-agnostic speech representations from multiple self-supervised
  tasks,'' \emph{arXiv preprint arXiv:1904.03416}, 2019.

\bibitem{shum2011exploiting}
S.~Shum, N.~Dehak, E.~Chuangsuwanich, D.~Reynolds, and J.~Glass, ``Exploiting
  intra-conversation variability for speaker diarization,'' in \emph{Proc. of
  Interspeech}, 2011, pp. 945--948.

\bibitem{narayanaswamy2019designing}
V.~S. Narayanaswamy, J.~J. Thiagarajan, H.~Song, and A.~Spanias, ``Designing an
  effective metric learning pipeline for speaker diarization,'' in \emph{IEEE
  International Conference on Acoustics, Speech and Signal Processing
  (ICASSP)}, 2019, pp. 5806--5810.

\bibitem{Filho2018}
V.~A. {Miasato Filho}, D.~A. Silva, and L.~G. {Depra Cuozzo}, ``Joint
  discriminative embedding learning, speech activity and overlap detection for
  the dihard speaker diarization challenge,'' in \emph{Proc. of Interspeech},
  2018, pp. 2818--2822.

\bibitem{diez2018speaker}
M.~Diez, L.~Burget, and P.~Matejka, ``Speaker diarization based on bayesian hmm
  with eigenvoice priors.'' in \emph{Odyssey}, 2018, pp. 147--154.

\bibitem{singh2019leap}
P.~Singh, H.~Vardhan, S.~Ganapathy, and A.~Kanagasundaram, ``{LEAP} diarization
  system for the second dihard challenge,'' in \emph{Proc. of Interspeech},
  2019, pp. 983--987.

\bibitem{zhang2019fully}
A.~Zhang, Q.~Wang, Z.~Zhu, J.~Paisley, and C.~Wang, ``Fully supervised speaker
  diarization,'' in \emph{IEEE International Conference on Acoustics, Speech
  and Signal Processing (ICASSP)}, 2019, pp. 6301--6305.

\bibitem{yang2016joint}
J.~Yang, D.~Parikh, and D.~Batra, ``Joint unsupervised learning of deep
  representations and image clusters,'' in \emph{Proceedings of the IEEE
  conference on computer vision and pattern recognition}, 2016, pp. 5147--5156.

\bibitem{kingma2014adam}
D.~P. Kingma and J.~Ba, ``Adam: A method for stochastic optimization,''
  \emph{arXiv preprint arXiv:1412.6980}, 2014.

\bibitem{wang2018speaker}
Q.~Wang, C.~Downey, L.~Wan, P.~A. Mansfield, and I.~L. Moreno, ``Speaker
  diarization with lstm,'' in \emph{IEEE International Conference on Acoustics,
  Speech and Signal Processing (ICASSP)}, 2018, pp. 5239--5243.

\end{thebibliography}
\end{document}